\documentclass[english,12pt]{article}

\usepackage[T1]{fontenc}
\usepackage[latin9]{inputenc}
\usepackage{amsmath}
\usepackage{amssymb}
\usepackage{bbm}
\usepackage{cancel}
\usepackage{subcaption}
\usepackage{caption}
\usepackage{comment}

\makeatletter

\usepackage{slashed}
\usepackage{physics}
\usepackage{tikz-cd}

\usepackage{hyperref}
\hypersetup{
	colorlinks,
	citecolor=blue,
	filecolor=blue,
	linkcolor=blue,
	urlcolor=blue,
	linktocpage=true
}

\makeatother

\usepackage[backend=bibtex,sorting=none,style=numeric-comp,url=false,maxbibnames=4]{biblatex}
\renewbibmacro{in:}{}
\usepackage{babel}
\bibliography{refs}{}

\usepackage{booktabs}
\newcommand{\ra}[1]{\renewcommand{\arraystretch}{#1}}

\usepackage{wrapfig}

\usepackage[left=2.5cm, right=2.5cm, top=3cm, bottom=3cm]{geometry}

\newcommand{\lb}{ \left( }
\newcommand{\rb}{ \right) }

\usepackage{pdfpages}
\usepackage{autobreak}
\usepackage{authblk}

\numberwithin{equation}{section}

\begin{document}
\title{More on chaos at weak coupling}
\author[1]{Rohit R. Kalloor}
\author[2]{Adar Sharon}
\affil[1]{Department of Particle Physics and Astrophysics, Weizmann Institute of Science, Rehovot, Israel}
\affil[2]{Simons Center for Geometry and Physics, SUNY, Stony Brook, NY 11794, U.S.A.}

\newcommand\mc[1]{\mathcal{#1}}
\newcommand\J{\mathcal{J}}
\newcommand\todo[1]{\textcolor{red}{(#1)}}
\newcommand\beq{\begin{equation}}
\newcommand\eeq{\end{equation}}
\newcommand\N{{\mathcal{N}}}

\setcounter{tocdepth}{2}

\maketitle

\begin{abstract}
We discuss aspects of the quantum Lyapunov exponent $\lambda_L$ in theories with an exactly marginal SYK-like random interaction, where $\lambda_L$ can be computed as a continuous function of the interaction strength $\mathcal{J}$. In $1d$, we prove a conjecture from \cite{Berkooz:2021ehv} which states that at small $\J$, $\lambda_L$ can be found by considering a specific limit of the four-point function in the decoupled theory. We then provide additional evidence for the $2d$ version of this conjecture by discussing new examples of Lyapunov exponents which can be computed at weak coupling. 
\end{abstract}

\newpage

\tableofcontents{}

\newpage

\section{Introduction}

The quantum Lyapunov exponent $\lambda_L$ is an intriguing and complicated observable in quantum field theories. Of particular interest are theories where $\lambda_L$ saturates an upper bound \cite{Maldacena:2015waa}, since this might indicate that they have semiclassical gravity duals. However, in this paper we will be interested in the opposite limit, and would like to study how the chaos exponent behaves as we turn on some small coupling $\J$. The classical version of this question has been studied extensively, and many interesting behaviors have been found for chaos at weak coupling. For example, the KAM theorem schematically states that an integrable system deformed by a small (integrability-breaking) deformation remains non-chaotic even for a finite but small enough deformation. We will ask an analogous question in a quantum setting: starting with $N$ decoupled theories and slowly increasing the strength of a random interaction between them, how does $\lambda_L$ behave?

Our setup is the following. The basic building block is a conformal field theory (CFT) $\mathcal{C}$, called the ``core CFT'', which contains a primary $\Phi$. Next, we take $N$ copies of the core CFT, and deform it by a random interaction:
\begin{equation}\label{eq:disordered_CFTs}
    \mathcal{C}^N+J_{i_1...i_q}\int d^dx\Phi_{i_1}...\Phi_{i_q}\;.
\end{equation}
The couplings $J_{i_1...i_q}$ are random with Gaussian measure and variance $\langle J_{i_1...i_q}^2\rangle=(q-1)!\frac{\J^2}{N^{q-1}}$ (with no sum over repeated indices).\footnote{We emphasize that our disorder is spacetime-independent.} 
The deformation should be understood in terms of conformal perturbation theory around $N$ copies of the CFT $\mathcal{C}$, and we will be studying these theories in the large-$N$ limit. We will call such theories ``disordered CFTs''. 

Disordered CFTs are a generalization of ideas appearing in disordered free theories, which can be obtained by setting $\mathcal{C}$ to be a free field theory, and setting $\Phi$ to be the corresponding free field. A famous example of disordered free theories is the SYK model \cite{PhysRevLett.70.3339,KitaevTalk}, and some additional examples can be found in \cite{Murugan:2017eto,Bulycheva:2018qcp,Fu:2016vas,Peng:2018zap,Chang:2021fmd,Chang:2021wbx,Gross:2017vhb,Lian:2019axs,Popov:2019nja}. In general, the random interactions allow for some exact computations in the IR for disordered free fields, assuming the theory flows to an interacting CFT \cite{PhysRevLett.70.3339,Maldacena:2016hyu,Kitaev:2017awl}. This was generalized to arbitrary core CFTs in \cite{Berkooz:2021ehv,Berkooz:2022dfr}. In these theories, $\lambda_L$ can be read off from a certain out-of-time-ordered four-point function (or alternatively, the double-commutator) \cite{KitaevTalk,KitaevTalk2,Larkin1969QuasiclassicalMI,Maldacena:2015waa}. This is a difficult computation in general, but it is aided by the large-$N$ limit and conformal invariance in the IR. In particular, we must also restrict ourselves to dimensions $d\leq 2$, since we will be interested in the theory at finite temperature, but in small dimensions a conformal transformation can map such theories to flat space.

An especially interesting subsector of disordered CFTs can be obtained by demanding that the interaction term in \eqref{eq:disordered_CFTs} is exactly marginal, such that $\mathcal{J}$ parametrizes a line of CFTs. This behavior is not generic, but can be obtained by using supersymmetry or by considering chiral theories, both of which we will discuss in this paper.\footnote{Similar behavior can be obtained in QM without these assumptions \cite{Gross:2017vhb}.} Having a line of CFTs allows us to follow interesting observables as we continuously move between different CFTs, and in particular will allow us to study $\lambda_L$ as a function of $\J$.

In \cite{Berkooz:2021ehv}, the chaos exponent was studied in the weak-coupling limit $\J\to 0$ in disordered CFTs where $\J$ is exactly marginal. In particular, in the examples that were discussed it was observed the chaos exponent in the limit $\J\to 0$ is also given by the leading exponential behavior of the double-commutator (DC) in a single core CFT. In other words, if the DC of a single core CFT behaves at large times as $\exp(\lambda_L^0 t)$, then the chaos exponent of the interacting theory in the limit $\J\to 0$ is given by 
\begin{equation}\label{eq:intro_continuity_temp}
    \lambda_L(J\to 0)=\lambda_L^0\;.
\end{equation}
It was conjectured that this result is general; we will call it the continuity conjecture.

\begin{figure}
	\centering
	\begin{subfigure}[t]{0.5\textwidth}
		\centering
		\includegraphics[width=0.5\linewidth]{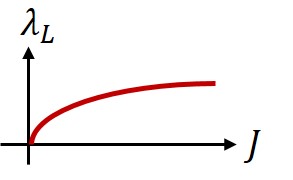}
		\caption{}
		\label{fig:continuous_chaos}
	\end{subfigure}
	~ 
	\begin{subfigure}[t]{0.5\textwidth}
		\centering
		\includegraphics[width=0.5\linewidth]{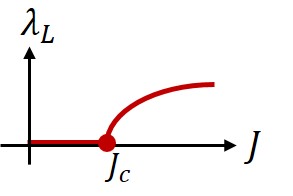}
		\caption{}
		\label{fig:discontinuous_chaos}
	\end{subfigure}
	\caption{Two types of behaviors for the dependence of $\lambda_L$ on the exactly marginal interaction $\J$: (a) continuous and (b) discontinuous.}
	\label{fig:dis_continuous_chaos}
\end{figure}

The equality \eqref{eq:intro_continuity_temp} is surprising for two main reasons. First, $\lambda_L^0$ cannot be interpreted as a chaos exponent in a single core CFT (since some form of a large-$N$ limit is required for this interpretation), and yet in the interacting theory it dictates the behavior of the chaos exponent at weak coupling. Second, it is possible for $\lambda_L^0$ to be \emph{negative}. A negative chaos exponent seems counter-intuitive, and indeed the result cannot be trusted in the standard method of computing $\lambda_L$ due to some assumptions that are required. A more precise version of the continuity conjecture is then
\begin{equation}\label{eq:intro_continuity}
    \lambda_L(J\to 0)=\max(\lambda_L^0,0)\;.
\end{equation}
Despite this fact, if $\lambda_L^0$ is negative then one can still determine that the chaos exponent of the theory is at most zero for a finite range of values of $\J$.
As a result, it is enough to show that in a single core CFT $\lambda_L^0<0$ in order to find a discontinuous transition into chaos, see figure \ref{fig:dis_continuous_chaos}. This is reminiscent of classical KAM theory. 

This paper includes two main results. First we will prove the continuity conjecture \eqref{eq:intro_continuity} in quantum mechanics (QM). We then discuss two examples in $2d$ and show that they obey the continuity conjecture, providing further evidence for the $2d$ version of the conjecture. The first example is the chiral SYK model discussed in \cite{Lian:2019axs}, where the chaos exponent was already found for all $\J$. We compare the result at small $\J$ to $\lambda_L^0$ and find agreement. Next we discuss the disordered $\mathcal{A}_3$ minimal model. Since this theory is dual to a free CFT, we can compute all $n$-point functions for its primaries, and we use this result to compute the chaos exponent at small $\J$. Comparing the result to $\lambda_L^0$ we again find agreement. 

In the examples discussed in this paper, $\lambda_L^0$ always turns out to be non-negative, so that the physical picture is that of figure \ref{fig:continuous_chaos}. We will discuss ideas for how to generate cases with a negative value and a discontinuous transition into chaos.

This paper is organized as follows. In section \ref{sec:background} we introduce the theories we consider in this paper and the methods for computing the chaos exponent. We then prove the continuity conjecture in QM in section \ref{sec:proof_QM}. In section \ref{sec:chiral_SYK} we discuss our first $2d$ example, the chiral SYK model, and show that the continuity conjecture is obeyed. In section \ref{sec:disordered_A3} we discuss our second example, the disordered $\mathcal{A}_3$ minimal model. We compute the chaos exponent at small $\J$ and again show that the continuity conjecture is obeyed. We discuss some future directions in section \ref{sec:conclusions}.

\section{Background}\label{sec:background}

In this section we review the basic method discussed in \cite{Berkooz:2021ehv} for obtaining the chaos exponent $\lambda_L$ at weak coupling for a general disordered CFT.

\subsection{Four-point function and double-commutator}

In disordered free theories, it is possible to write down self-consistency equations for the two- and four-point functions of the fundamental field, which can then be solved using a conformal ansatz when the conformal symmetry is restored in the IR. In \cite{Berkooz:2021ehv} it was shown that similar equations can be written down for general disordered CFT for the operator $\Phi$ in \eqref{eq:disordered_CFTs}, which can in principle be solved for any value of $\mathcal{J}$ when it is an exactly marginal deformation. However, solving these equations requires knowledge of all $n$-point functions of $\Phi$ in the core CFT, and so is very difficult in general. In this section we will review the derivation in \cite{Berkooz:2021ehv}, omitting some details. 

In the following we will only use the four-point function, and so we only discuss the self-consistency equation for the four-point function and for the DC. The discussion can also be immediately generalized to SUSY theories, in which case the diagrams discussed should be understood as supergraphs.

\begin{figure}[]
	\centering
	\includegraphics[width=0.75\linewidth]{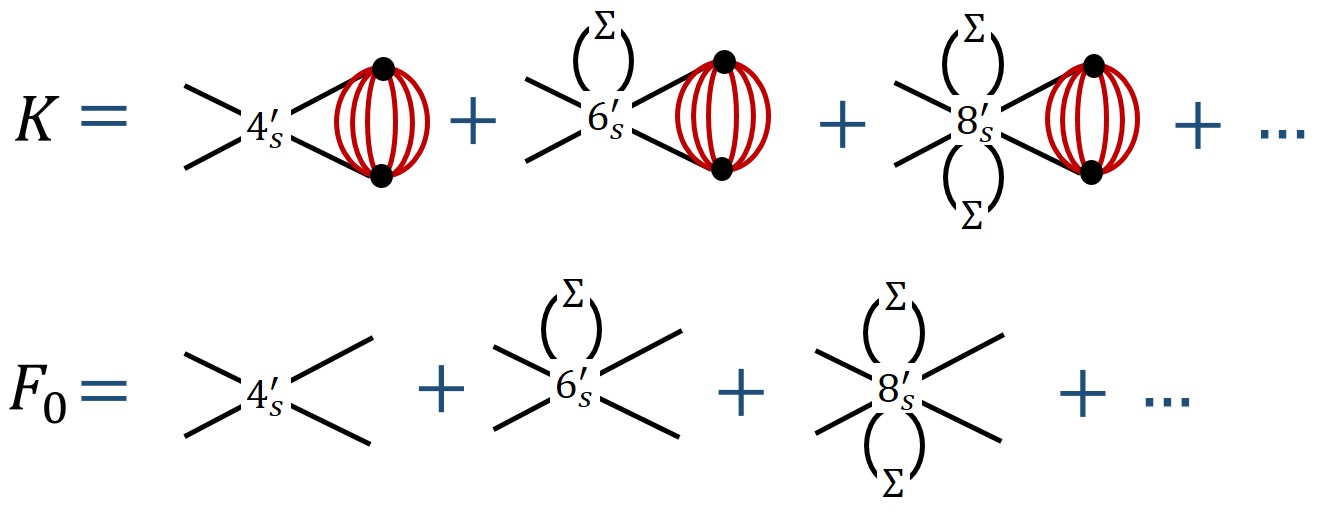}
	\caption{The kernel $K$ and initial diagram $F_0$ for general disordered CFTs. Red lines denote full propagators $G$, and black dots denote insertions of the disorder interaction, with $q-2$ red propagators between each pair.}
	\label{fig:kernel}
\end{figure}

\begin{figure}[]
	\centering
	\includegraphics[width=0.7\linewidth]{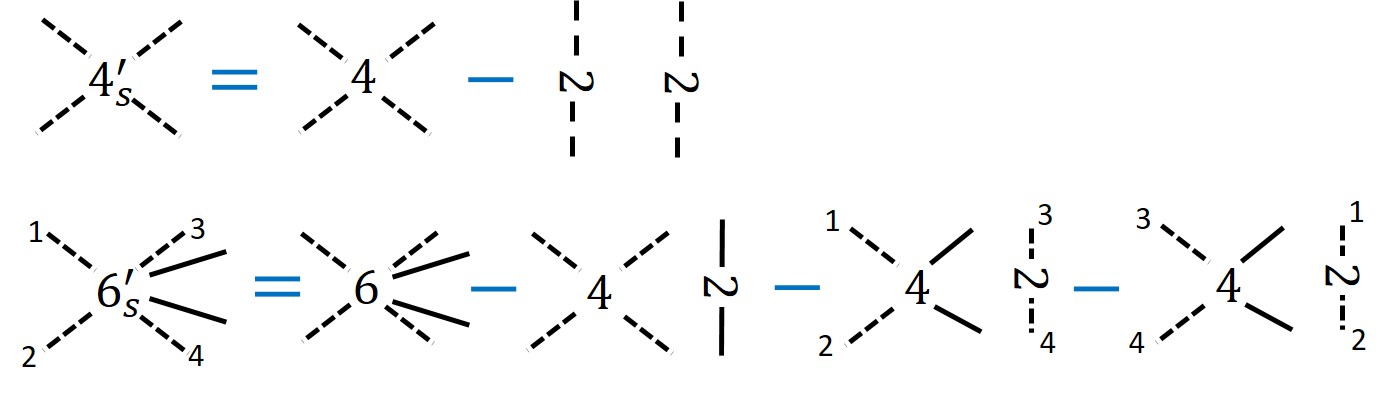}
	\caption{Examples of correlation functions $n_s'$. Dashed lines corresponds to external points, while solid lines are connected via $\Sigma$'s in figure \ref{fig:kernel}.}
	\label{fig:nsprime_examples}
\end{figure}

We would first like to compute the connected four-point function
\begin{equation}\label{eq:conn_four}
C=\frac{1}{N^2} \sum_{i,j=1}^N\langle\Phi_i\Phi_i\Phi_j\Phi_j\rangle_{conn}
\end{equation}
in the deformed theory \eqref{eq:disordered_CFTs}. The diagrams contributing to $C$ have an iterative ladder structure at large $N$, so that the full result for the four-point function is given by
\begin{equation}\label{eq:C_K_F0}
C=\sum_{n=0}^\infty K^n F_0=\frac{F_0}{1-K}\;,
\end{equation}
where $F_0,K$ are defined in figure \ref{fig:kernel}.
The computation of $K$ and $F_0$ requires knowing all ``subtracted $n$-point functions'', denoted by $n_s'$. These are given by the standard $n$-point functions but with some specific subtractions of lower-order correlators, and are explicitly defined in \cite{Berkooz:2021ehv}. Examples of some subtracted $n$-point functions appear in figure \ref{fig:nsprime_examples}. From these it is in principle possible to compute the full four-point function $C$. One can also perform perturbation theory in $\J$; at leading order only the first diagram in the expression for $K$ in figure \ref{fig:kernel} contributes, which is determined by the CFT four-point function and two-point function, and takes the form
\begin{equation}
    K(1,2;3,4)=(q-1)J^2 \langle \Phi_1\bar\Phi_2\Phi_3\bar\Phi_4 \rangle_s' G(3,4)^{q-2}\;,
\end{equation}
with $G(3,4)$ the $\Phi$ propagator between the points $x_3,x_4$.

Next we discuss the computation of the double-commutator, defined as
\begin{equation}\label{eq:OTOC_O}
\begin{split}
W_R(t_1,t_2) &= \frac{1}{N^2}
\sum_{i,j=1}^N \left\langle
[\Phi_i(\beta/2),\Phi_j(\beta/2+i t_2)] [\Phi_i(0),\Phi_j(i t_1)]
 \right\rangle\\
&=\lim_{\varepsilon\rightarrow0}\frac{1}{N^2}
\sum_{i,j=1}^N \left\langle \left( 
\Phi_i\left(\varepsilon\right)-\Phi_i\left(-\varepsilon\right) \right) 
\left(\Phi_i\left(\beta/2+\varepsilon\right)-\Phi_i\left(\beta/2-\varepsilon\right) \right)
\right.\\
& \qquad\qquad\qquad\qquad\qquad\qquad\qquad\qquad\qquad \cdot \left.
\Phi_j\left(i t_1\right)
\Phi_j\left(\beta/2+i t_2\right) 
\right\rangle\;.
\end{split}
\end{equation}
We have suppressed the spatial coordinates, since they will be unimportant, and we have kept the (real and imaginary) time coordinates. By $\langle ... \rangle$ we mean the Euclidean time-ordered thermal trace. Note that \eqref{eq:OTOC_O} is just a combination of analytically-continued Euclidean four-point functions on the cylinder, which in a $d=2$ CFT is an analytically-continued flat-space correlator \eqref{eq:conn_four}. 

\begin{figure}[]
	\centering
	\includegraphics[width=0.5\linewidth]{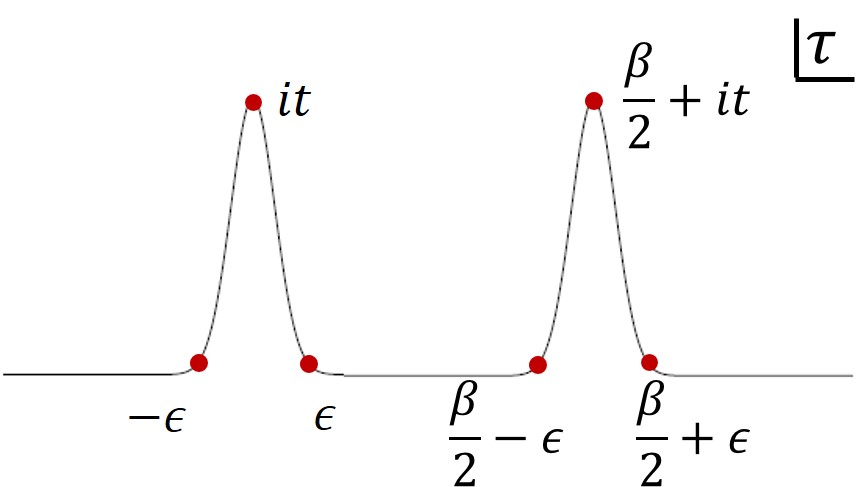}
	\caption{The complex time contour chosen for the computation of the DC. Red dots denote insertion points of operators. We call the two excursions from the real axis the ``left rail'' and the ``right rail''.}
	\label{fig:time_contour}
\end{figure}

The diagrams contributing to the double-commutator also obey an iterative ladder structure. The corresponding kernel $K_R$ and initial diagram $F_{0,R}$ have the same diagrammatics as the four-point function (appearing in figure \ref{fig:kernel}), but the correlators are analytically-continued versions of the ones appearing above. The details appear in \cite{Berkooz:2021ehv}, and we will write down explicitly only the leading contribution in $\J$, which comes from the four-point function:
\begin{equation}\label{eq:leading_kernel}
    K_R(t_1,t_2,t_3,t_4) = \J^2 \left\langle \Delta\mathcal{O}\left(it_1 \right)\Delta\mathcal{O}\left(\frac{\beta}{2}+it_{2}\right)\mathcal{O}\left(it_3 \right)\mathcal{O}\left(\frac{\beta}{2}+it_4 \right)\right\rangle^\prime_{s} G^{q-2}_{lr,\Delta}(3,4)+O(\J^4)\;.
\end{equation}
Here we have denoted $\Delta\mathcal{O}(z)=\mathcal{O}(z+\varepsilon)-\mathcal{O}(z-\varepsilon)$ where we eventually take the limit $\epsilon\to 0$. The integration range in principle for all points is over the complex time contour appearing in figure \ref{fig:time_contour}, although various cancellations as we take $\epsilon\to 0$ lead to the result above. In particular, we have used the expression for the thermal two-point function for a scalar operator of dimension $\Delta$ between points from different ``rails'' (see figure \ref{fig:time_contour}):
\begin{equation}\label{eq:Glr}
    G_{lr,\Delta}(1,2) = \frac{1}{\left( 4\cosh(\frac{t_{12}-x_{12}}{2})\cosh(\frac{t_{12}+x_{12}}{2}) \right)^\Delta}\;.
\end{equation}

\subsection{Chaos}\label{sec:chaos}

Having written down the ladder diagrams which contribute to the DC, we can now compute the chaos exponent. In principle, this is done by computing the DC explicitly, and then taking the large-time limit $t_1,t_2=t\to \infty$, which should lead to exponentially growing behavior,
\begin{equation}\label{eq:guess}
    W_R(t_1,t_2)\sim \exp(\frac{\lambda_L}{2}(t_1+t_2))f(t_1-t_2)\;.
\end{equation}
where from now on we set $\beta=2\pi$. However, computing the full DC is difficult, and instead the existence of an iterative ladder structure offers us a shortcut. The ladder structure leads to the self-consistency equation
\begin{equation}\label{eq:4pt_Self_consistency}
    W_R=F_{0,R}+K_R W_R\;.
\end{equation}
At large times, we assume that the $F_{0,R}$ term is negligible, and so $W_R$ obeys the equation
\begin{equation}
    W_R=K_R W_R\;,
\end{equation}
which is just an eigenvalue equation for $K_R$. As a result, the exponentially-growing solution for $W_R$ must be an eigenfunction of the retarded kernel $K_R$ with eigenvalue $1$. Thus, the chaos exponent is found by guessing solutions of the form \eqref{eq:guess} and finding their eigenvalue $k_R(\lambda_L)$ under $K_R$. The largest $\lambda_L$ for which $k_R(\lambda_L)=1$ is the chaos exponent. 

The precise form of the eigenvalue is constrained due to conformal invariance, and takes the form of a two-point function on the cylinder. In $2d$ we have the ansatz
\begin{equation}\label{eq:W_ansatz}
    W_\lambda(1,2) = \frac{\exp(-\frac{h+\tilde h}{2}(t_1+t_2)+\frac{h-\tilde h}{2}(x_1+x_2))}
    {(2\cosh(\frac{t_{12}-x_{12}}{2}))^{\Delta-h}
    (2\cosh(\frac{t_{12}+x_{12}}{2}))^{\Delta-\tilde h}}\;.
\end{equation}
In general we require $h=-\frac{\lambda}{2} + i p, \tilde h=-\frac{\lambda}{2} - i p$ for real $\lambda, p$, but in practice in the examples appearing here the maximal chaos exponent will have $p=0$.

Finding the chaos exponent now amounts to computing the eigenvalue of the eigenfunction $W_\lambda$ under $K_R$. Since the theories we consider are conformally invariant for any value of $\J$, we can perform this computation in a perturbative expansion in $\J$. Generally the eigenvalue is given by
\begin{equation}\label{eq:kR}
    k_{R}(\lambda,\J)  = \frac{\int d^2 x_3 d^2 x_4 K_R \cdot W}{W}\;,
\end{equation}
and plugging in the leading-order result for $K_R$ we find that in $2d$ and at leading order in $\J$,
\begin{equation}\label{eq:kR_4p}
\begin{split}
    k_R(\lambda,\J)
    =& \J^2 \int d^2 x_3 d^2 x_4 \left\langle \Delta\mathcal{O}\left(it_1 \right)\Delta\mathcal{O}\left(\beta/2+it_{2}\right)\mathcal{O}\left(it_3 \right)\mathcal{O}\left(\beta/2+it_4 \right) \right\rangle'_s \\
    & \cdot \frac{G_{lr,\Delta+\frac{\lambda}{2}}(3,4)}{G_{lr,\Delta+\frac{\lambda}{2}}(1,2)} \exp\left(\frac{\lambda}{2} (t_3+t_4-t_1-t_2)\right)
    \cdot G_{lr,2\Delta(q-2)}(3,4) +O(J^4)\\
      =& \frac{\J^2}{4} \frac{\exp(-\frac{\lambda}{2} (t_1+t_2))}{G_{lr,\lambda/2}(1,2)} 
     \frac{\int^\infty_{u_1} du_3 \int_{-\infty}^{u_2} du_4}{u_{34}^{2+\frac{\lambda}{2}}}
     \frac{\int_{v_1}^\infty dv_3 \int_{-\infty}^{v_2} dv_4}{v_{34}^{2+\frac{\lambda}{2}}} \mathcal{G}_R(\chi,\bar\chi)  +O(\J^4)\;.
\end{split}
\end{equation}
With $\chi,\bar\chi$ the conformal cross-ratios. Here we have performed the change of variables
\begin{equation}
    \begin{split}
        u_3=e^{x_3-t_3}\;,&\quad v_3=e^{-x_3-t_3}\;,\\
        u_4=-e^{x_4-t_4}\;,&\quad v_4=-e^{-x_4-t_4}\;.
    \end{split}
\end{equation}
$\mathcal{G}_R$ is the retarded normalized 4-point of the undeformed CFT at $\J=0$,
\begin{equation}\label{eq:def_G_R}
\begin{split}
    \mathcal{G}_R(\chi,\bar\chi) &= \frac{\left\langle  
    [O(\beta/2+it_2),O(\beta/2+it_4)]
    [O(it_1),O(it_3)]
    \right\rangle_s'}{G_{lr,\Delta}(1,2)G_{lr,\Delta}(3,4)}\\
    & = \lim_{\varepsilon_1,\varepsilon_2\rightarrow 0} \mathcal{G}_{++}-\mathcal{G}_{+-}-\mathcal{G}_{-+}+\mathcal{G}_{--}
\end{split}
\end{equation}
with the normalized four-point $\mathcal{G}_{\pm_1,\pm_2} = \mathcal{G}(u_1 e^{\pm i \varepsilon_1}, v_1 e^{\pm i \varepsilon_1},u_2 e^{\pm i \varepsilon_2}, v_1 e^{\pm i \varepsilon_2},u_3,v_3,u_4,v_4)$.

\subsubsection{Consistency of the perturbative expansion}\label{sec:consistency_of_perturbative}

We now discuss perturbative corrections to the chaos exponent which are subleading in the $\J\to 0$ limit. The computation of the chaos exponent in perturbation theory in $\J$ requires several assumptions. Expanding the eigenvalues of the retarded kernel, we expect to see
\begin{equation}
    k_R(\lambda,\mathcal{J})=\mathcal{J}^2f_2(\lambda)+\mathcal{J}^4f_4(\lambda)+...\;,
\end{equation}
First, we would like to understand when it is enough to find  leading term $f_2(\lambda)$ discussed above in order to find the leading value of the chaos exponent in the limit $\mathcal{J}\to 0$. The chaos exponent is found by setting $k_R=1$, and so if we keep only the $\J^2$ term, $\lambda_L$ is found by analyzing at which values of $\lambda$ the function $f_2(\lambda)$ diverges as $1/\J^2$. The largest such $\lambda$ can then be identified with the chaos exponent. In order for this procedure to be consistent it is enough to require two conditions on the higher-order terms:
\begin{enumerate}
    \item If the largest value of $\lambda$ at which $f_2$ diverges is $\lambda_0$, then all other $f_n$ diverge at values $\lambda\leq \lambda_0$. 
    \item If $f_2$ diverges as $\frac{1}{(\lambda-\lambda_0)^\alpha}$, then other $f_n$ diverge as $\frac{1}{(\lambda-\lambda_0)^{\beta_n}}$ for $\beta_n\leq n\alpha/2$.
\end{enumerate}
In this case the leading value of $\lambda_L$ is indeed $\lambda_0$. All examples discussed in \cite{Berkooz:2021ehv} were shown to obey these conditions, and we will show that the examples discussed here also obey them).

Next, it is natural to ask when it is consistent to perform a perturbative expansion around $\lambda_0$ in the limit of small $\J$ in order to obtain the chaos exponent in a series in $\mathcal{J}$ of the form
\begin{equation}
    \lambda_L=\lambda_0+\mathcal{J}^2\lambda_1+...
\end{equation}
This requires a strict inequality in the second condition above, i.e.~we require
\begin{equation}
    \beta_n<n\alpha/2\;.
\end{equation}
To see why this is the case, assume an expansion of the form 
\begin{equation}
    k_{R}=\sum a_n \frac{\J^{2n}}{\lambda^{2n}}\;,    
\end{equation}
so that $\alpha=2$ and $\beta_n=n=n\alpha/2$ and the inequality is exactly saturated. As a result, the expansion is not in terms of $\J$, but in terms of $\J/\lambda$. We can now try to compute $\lambda_L$ in perturbation theory. Write
\begin{equation}
    \lambda_L=\lambda_0+\J^2 \lambda_1+...
\end{equation}
Plugging in this expansion, we immediately learn that the leading order is given by $\lambda_0=0$, as discussed above. However, in order to find the value of the subleading term $\lambda_1$, we must take into account all of the terms in the expansion, since all terms they are all of the same order in $\mathcal{J}$.
As a result, while we can still compute $\lambda_0$ in such theories, we cannot perform perturbation theory around it without knowing $k_R$ completely (or at least to all orders in some double-scaling limit). So a perturbative calculation of $\lambda_L$ beyond the leading order using this method fails. We will see this happen in the chiral SYK example discussed in section \ref{sec:chiral_SYK}. 

On the other hand, for any theory where there is a strict inequality $\beta_n<n\alpha/2$, a perturbative expansion in $\mathcal{J}$ should be possible. This is the result in e.g. the generalized free fields examples in \cite{Berkooz:2021ehv}.

\subsection{The continuity conjecture}\label{sec: continuity}

We now discuss the behavior of the chaos exponent as $\J\to 0$. Note that in equation \eqref{eq:kR_4p}, the prefactor $\J^2$ vanishes in this limit, and so in order for us to be able to solve for $k_R(\lambda_L,\J)=1$, we must find values of $\lambda$ for which the integral in \eqref{eq:kR_4p} diverges as $1/\J^2$. The chaos exponent in this limit is then given by the largest value of $\lambda$ for which this divergence occurs. Assuming that the retarded four-point function of the theory at a single core CFT behaves at large times $t_1,t_2$ as\footnote{In $d\leq 2$ this is related to the Regge limit.}
\begin{equation}\label{eq:large_times}
    \mathcal{G}_R(t_1,t_2)\sim \exp(\lambda_L^0(t_1+t_2)/2)\;,
\end{equation}
it is easily seen that the divergence occurs precisely at $\lambda=\lambda_L^0$. As a result, it was conjectured in \cite{Berkooz:2021ehv} that the chaos exponent in the limit $\J\to 0$ is given precisely by the leading exponential behavior of the double-commutator in the free theory at $\J=0$. Thus the computation of the leading behavior of $\lambda_L$ is particularly simple, and is a property of a single core CFT.\footnote{Note that $\lambda_L^0$ does not have any interpretation in terms of a chaos exponent in a single core CFT, since this requires some form of a large-$N$ limit or another weak-coupling expansion.} Several examples were discussed in \cite{Berkooz:2021ehv} which were shown to obey this conjecture.

\section{Proving the continuity conjecture in QM}\label{sec:proof_QM}

The continuity conjecture discussed in \ref{sec: continuity} relates the chaos exponent in disordered CFTs at small $\mathcal{J}$ to the late-time behavior of the DC in a single core CFT. It states that assuming that $\mathcal{G}_R$ defined in \eqref{eq:def_G_R} (or equivalently, the double-commutator) behaves at large times as
$
  \exp(\lambda_L^0 t)
$,
then the chaos exponent as $\J\to 0$ approaches $\lambda_L^0$. We will now prove this statement under the assumption that perturbation theory in $\J$ is valid for the eigenvalues of the retarded kernel, so that the leading order is obtained by considering just the contribution of the four-point function to the retarded kernel, see section \ref{sec:chaos}.

Consider the leading contribution to $k_R$, which comes from the four-point function diagram in figure \ref{fig:kernel}. In $1d$ this contribution simplifies to (see Appendix \ref{app:1d_integral} for a derivation):

\begin{equation}\label{eq:1d_integral}
k_R =\frac{1}{|z_{12}|^{2\Delta}}\int^\infty_{z_1} dz_3  \int^{z_2}_{-\infty} dz_4  \frac{\mathcal{G}_R(\chi)}{|z_{34}|^{2+\lambda}}\;.
\end{equation}
Here, we performed the change of variables 
$z=e^{-t}$ on the ``left rail'' (points $1,3$) and $z=-e^{-t}$ on the ``right rail'' (points $2,4$), see e.g.~\cite{Murugan:2017eto} for details. The conformal cross-ratio is 
\begin{equation}
    \chi=\frac{z_{12}z_{34}}{z_{13}z_{24}}\;.
\end{equation}

Next we perform another change of coordinates to the coordinates $\kappa,\chi$, where $\kappa$ is defined as $z_3=\kappa/\chi$. 
The integral becomes is
\begin{equation}
    k_R(\lambda) = \int_{-\infty}^{0} d\chi\frac{\mathcal{G}_R(\chi)}{\chi^{1-\lambda}} \int_{-\infty}^{\chi-1}  \frac{d\kappa}{\kappa^{1+\lambda}(\kappa-\chi)^{1+\lambda}(\kappa-(\chi-1))^{-\lambda}}\;.
\end{equation}
The chaos exponent $\lambda_L$ is now given by the largest value of $\lambda$ for which the integral diverges in the limit $t_3,t_4\to -\infty$, which in these coordinates corresponds to $\chi \rightarrow 0$.
Note that the $\kappa$ integral is finite as long as $\lambda>-1$ (corresponding to $\lambda_L^0>-1$). At $\chi=0$ the $\kappa$ integral is smooth; it does not vanish or diverge. Therefore we can expand $\mathcal{G}_R(\chi)$ around $
\chi=0$ inside the integral. Let us denote the leading order term by $\mathcal{G}_R(\chi)=c_0 \chi^{-\lambda_L^0}+...$. Note that since at large times we have $\chi\sim e^{-t}$, we can identify $\lambda_L^0$ with the rate of growth of the DC of a single core CFT as in \eqref{eq:large_times}. Plugging in this expansion, we see that the integral converges only for $\lambda>\lambda_L^0$, which means that the chaos exponent at small $\J$ is $\lambda_L^0$. We have thus  proved the continuity conjecture in $1d$.

\section{The chiral SYK model}\label{sec:chiral_SYK}

The chiral SYK model was introduced in \cite{Lian:2019axs} (see also \cite{Hu:2021hsj}). The theory is a disordered free theory in $2d$, where the core CFT is a free chiral Majorana fermion. The theory was shown to have a line of fixed points characterized by $\mathcal{J}$, so that the chaos exponent can be found as a function of $\mathcal{J}$. Since the theory is chiral, instead of the standard chaos exponent it is more interesting to consider the velocity-dependent chaos exponent, given by considering a large time and large distance limit with the ratio $v=x/t$ kept constant. The corresponding chaos exponent is denoted $\lambda_v$. It was found that the chaos exponent always starts at zero as $\mathcal{J}\to 0$, and rises as we increase $\mathcal{J}$. In particular, choosing $v$ such that $\lambda_v$ is maximal, one finds that at infinite $\mathcal{J}$ the chaos exponent $\lambda_v$ saturates the bound on chaos \cite{Maldacena:2015waa}. We will be interested in the weak-coupling limit, where $\mathcal{J}$ is close to zero. We will show that the continuity conjecture is obeyed for the velocity-dependent chaos exponent, and discuss corrections in $\mathcal{J}$.

The velocity-dependent chaos exponent at weak coupling can be easily extracted from Appendix B of \cite{Lian:2019axs} (see equation (B.6)), and in the limit $\J\to 0$ one finds
\begin{equation}
    \lambda_v(\J)=\begin{cases} \frac{2\pi}{\beta}\eta(v-1)+O(\J),& u_-<v<u_+\\
    0,& \text{else}
    \end{cases}
\end{equation}
where $u_\pm=1\pm \frac{\mathcal{J}}{2\pi}$ and
\begin{equation}
    \eta = \frac{\sqrt 3}{\sqrt{1-\J^2}}\frac{1-v}{\sqrt {\J^2-(1-v)^2}}\;.
\end{equation}
Note that since $u_-<v<u_+$, $\eta$ is finite in the limit $\J\to 0$. 

Since the theory is chiral, it is not surprising that the chaos exponent vanishes outside of a cone around the speed of light. Taking the strict $\J\to 0$ limit, we find that $\lambda_v(\J\to 0)$ vanishes trivially unless $v=1$ (since $u_-=u_+=1$), but at $v=1$ it turns out to vanish as well. Thus we find $\lambda_v(\J\to 0)=0$.

We would like to compare this to $\lambda_L^0$, which describes the large-time behavior of the DC in a single copy of the free core CFT. The DC in the core CFT is given by $G_R(13)G_R(24)$, where $G_R(ij)$ is the retarded propagator between spacetime points $i,j$ and is given by
\begin{equation}
G_{R}(t, x)=\frac{1}{\beta \sqrt{u_{+} u_{-}}} \frac{\Theta\left(t-u_{+}^{-1} x\right) \Theta\left(u_{-}^{-1} x-t\right)}{\sqrt{\sinh \left[\frac{\pi}{\beta}\left(t-u_{+}^{-1} x\right)\right] \sinh \left[\frac{\pi}{\beta}\left(u_{-}^{-1} x-t\right)\right]}}\;.
\end{equation}
Next we take the large-time limit where $t=t_1=t_2$ and $x=x_1=x_2$ are large while $v=x/t$ is kept constant. We find (up to unimportant constants)
\begin{equation}
    DC\propto
    \begin{cases}\exp\left(-\frac{\pi}{\beta}(u_-^{-1}-u_+^{-1})vt\right) ,&u_-<v<u_+\\
    0,&\text{else}\end{cases}
\end{equation}
Plugging in $\mathcal{J}=0$ we find as a result that $\lambda_v^0=0$ always. Thus we have found
\begin{equation}
    \lambda_v^0=\lambda_v(\J\to 0)=0\;,
\end{equation}
and so the continuity conjecture is obeyed.

We can also try to understand subleading corrections to the chaos exponent as discussed in section \ref{sec:consistency_of_perturbative}. The eigenvalues of the chiral SYK model were computed in \cite{Lian:2019axs}, and we can expand the result in $\mathcal{J}$:
\begin{equation}
    k_{R}=\frac{3\mathcal{J}^2}{\lambda^2}-\frac{6\mathcal{J}^3}{\lambda^3}+O\left(\mathcal{J}^4\right)\;.    
\end{equation}
Expanding to higher orders one finds $\alpha=2$ and $\beta_n=n=n\alpha/2$ in the notation of section \ref{sec:consistency_of_perturbative}. As a result, the expansion is not in terms of $\J$, but in terms of $\J/\lambda$, and so a perturbative computation of $\lambda_L$ will fail at higher orders in $\J$.

\section{The disordered \texorpdfstring{$\mathcal{N}=2$ } \texorpdfstring{$\mathcal{A}_3$} minimal model}\label{sec:disordered_A3}

We now discuss the case where the core CFT is the $\mathcal{N}=2$ supersymmetric $\mathcal{A}_3$ minimal model. This model can be constructed using a single chiral superfield $X$ with superpotential
\begin{equation}
    W=X^4\;.
\end{equation}
The model has central charge $c=3/2$ and its spectrum includes a chiral primary of dimension $1/4$ which we will call $\Phi$.
This theory can be identified with the theory of a free boson $H$ an a free fermion $\chi$, which combine into a single free $\mathcal{N}=1$ chiral multiplet \cite{Mussardo:1988av}. The core CFT thus has a free field representation.

The disordered $\mathcal{A}_3$ minimal model with $q=4$ can then be constructed as
\begin{equation}
    (\mathcal{A}_3)^N+\sum_{i_1\neq...\neq i_4}J_{i_1...i_4}\int d^2xd^2\theta \Phi_{i_1}...\Phi_{i_4}\;.
\end{equation}
In particular, this deformation is marginal. In fact, as discussed in \cite{Berkooz:2021ehv}, it can be shown using standard arguments \cite{Green:2010da,Kol:2002zt,Kol:2010ub} that any realization of this theory is a CFT, so that the interaction is exactly marginal (even without averaging).

In order to compute $\lambda_L$ we need to know $n$-point functions of $\Phi$. Since the $\mathcal{A}_3$ minimal model has a free field realization in terms of $\mathcal{N}=1$ superfields, we should be able to identify the components of $\Phi$ with products of operators from the free boson and free fermion CFT. In practice, the various components will be mapped to products of vertex operators and fermionic twist fields, whose $n$-point functions are known. Plugging the results into the retarded kernel, we will be able to read off $\lambda_L$. Following the discussion above, we will focus on the four-point function which gives the leading contribution to $\lambda_L$ at small $\J$.

Our notations appear in appendix \ref{app:superappendix}. In particular, we use lightcone coordinates $x^\pm$ so that a two-point function takes the form $(x^+x^-)^{\Delta}\equiv |x|^{2\Delta}$.

\subsection{Details of duality to free fields}

The free field representation of the $\mathcal{A}_3$ minimal model consists of a free Majorana fermion $\chi$ and a free compact boson $H$ at the self-dual radius $R=1/\sqrt 2$ (see for instance \cite{Mussardo:1988av} or \cite{Polchinski:1998rr}). The SUSY algebra is generated by the operators
\begin{align}
    G _{ \pm } & = \chi _{\pm } \exp \lb i \sqrt{2} H_{\pm } \rb\;,
    \nonumber \\
    \bar{G} _{ \pm } & = \chi _{\pm } \exp \lb - i \sqrt{2} H_{\pm } \rb\;,
    \nonumber \\
    j_{\pm} ^{(R)} & = \frac{i }{\sqrt{2} } \partial _{\pm} H\;,
\end{align}
where $\pm $ stand for the left/right moving parts of the fields and operators.\footnote{The theory actually has $\mathcal{N}=3$ SUSY, but we will only need the $\mathcal{N}=2$ subalgebra above for our purposes.}

The fundamental superfield $\Phi $ has scaling dimensions $ \lb \tfrac{1}{8}, \tfrac{1}{8} \rb $, and there are five superprimaries in the NS sector: $\Phi $, $\Phi ^2$, $\bar{\Phi} $, $\bar{\Phi }^2 $, $\bar{\Phi }\Phi $. Their bottom components map to the following free theory operators:
\begin{align}
\label{eq:field_identification}
    \phi = \sigma \exp \lb i \frac{H}{ 2 \sqrt{2} } \rb\;,  & \qquad 
        \bar{\phi } = \sigma \exp \lb - i \frac{H}{ 2 \sqrt{2} } \rb\;,
    \nonumber \\
    \phi ^2 = \exp \lb i \frac{H}{ \sqrt{2} } \rb\;,  & \qquad 
        \bar{\phi } ^2 = \exp \lb - i \frac{H}{ \sqrt{2} } \rb\;,
    \nonumber \\
    \bar{\phi} \phi & = \epsilon = \chi _+ \chi _-\;,
\end{align}
where $\sigma $ and $\mu $ (which will be of use later), with $(h, \bar h) = (1/16,1/16)$, are the twist fields of the fermion theory. The other components of these superfields may be worked out via free field OPEs (see Appendix \ref{app:superappendix}). For convenience, we list in Table \ref{tab:freerep} the components of the basic superfield:
\begin{align}
    \Phi (y^+,y^-) = \phi(y^+, y^-)+\theta^+\psi_+(y^+, y^-)+\theta^-\psi_-(y^+, y^-)+\theta^+\theta^- F(y^+, y^-)
\end{align}
along with their dimensions and R-charges.

\begin{table}
\centering
\ra{1.5}
\caption{The operators in the multiplet of $ \Phi $ in the $\mathcal{A}_3 $ minimal model.}
\label{tab:freerep}
\begin{tabular}{@{}cllc@{}}
\toprule
            $\mathcal{O}$ & $\mathcal{O}_{\text{free}} $   &    $ ( h ,\bar{h} ) $ & $q_R $  \\
            \cmidrule{1-4}
            $ \phi $ & $ \sigma \ e^{i \frac{H}{2 \sqrt{2} } }  $ & $ \lb \frac{1}{8} , \frac{1}{8} \rb $ & $ \lb \frac{1}{4} , \frac{1}{4} \rb $ \\
            $ \psi _{+} $ & $ \mu \ e^{i \frac{-3 H_+ + H_{-} }{2 \sqrt{2} } }  $ & $ \lb \frac{5}{8} , \frac{1}{8}\rb $ & $ \lb -\frac{3}{4} , \frac{1}{4} \rb $ \\
            $ \psi _{-} $ & $ \mu \ e^{i \frac{ H_+ - 3H_{-} }{2 \sqrt{2} } }  $ & $ \lb \frac{1}{8} ,\frac{5}{8} \rb $ & $ \lb \frac{1}{4} , -\frac{3}{4} \rb $\\
            $ F $ & $ \sigma \ e^{- i \frac{3 H}{2 \sqrt{2} } }  $ &  $ \lb \frac{5}{8} ,\frac{5}{8} \rb $ & $ \lb -\frac{3}{4} , -\frac{3}{4} \rb $ \\
            \bottomrule
        \end{tabular}
\end{table}

\subsection{Correlation functions}

Having identified the components of the superfield $\Phi$ with operators from free field theories, we can now compute all $n$-point functions of the field $\Phi$.

\subsubsection{Two-point function}

Superconformal symmetry fixes the form of the two-point function to be 
\begin{equation}
    \langle \Phi \Bar\Phi\rangle = \frac{1}{|\langle12\rangle|^{2\Delta}}
\end{equation}
where the relevant value for our theory is $\Delta=1/4$.
By expanding the result in the superspace coordinates on both sides, we can identify two-point functions of the various components of $\Phi$, which allows us to set their normalization. We find
\begin{equation}\label{eq:props}
\begin{split}
    \langle \phi\bar\phi \rangle= \frac{1}{|x_{12}|^{2\Delta}}\;, &  \qquad \langle F\bar F \rangle= \frac{4\Delta^2}{|x_{12}|^{2(\Delta+1)}}\;, \\ 
    \langle \psi^+\bar\psi^+ \rangle= -\frac{2\Delta}{|x_{12}|^{2\Delta}x_{12}^+} \;, & \qquad \langle \psi^-\bar\psi^- \rangle=-\frac{2\Delta}{|x_{12}|^{2\Delta}x_{12}^-}\;.
\end{split}
\end{equation}

\subsubsection{Four-point function}

We move on to the computation of the four-point function. Superconformal symmetry fixes its form to be
\begin{align}
    \langle \Phi(x_1)\bar{\Phi}(x_2)\Phi(x_3)\bar{\Phi}(x_4) \rangle = \frac{1}{|\langle12\rangle\langle34\rangle|^{\frac{1}{2}}}f(\chi_s^+,\chi_s^-)\;,
\end{align}
where $f$ is an arbitrary function of the superconformal cross ratios
\begin{equation}
    \chi_s^\pm=\frac{\langle12\rangle^\pm\langle34\rangle^\pm}{\langle14\rangle^\pm\langle32\rangle^\pm}\;.
\end{equation}
Note that the bottom component of $\chi_s^\pm$ is the usual non-supersymmetric cross-ratio $\chi^\pm=\frac{x_{12}^\pm x_{34}^\pm}{x_{14}^\pm x_{32}^\pm}$.

We start by calculating the bottom component of this four point function. Using the identification \eqref{eq:field_identification}, this is equivalent to computing the product of a four-point function of vertex operators and of twist operators $\sigma$. The results are well known, and we find
\begin{equation}
    \langle 
    \phi(x_1)\bar{\phi}(x_2)\phi(x_3)\bar{\phi}(x_4) \rangle=
    \left|
    \frac{1}{x_{12}x_{34}}
    \right|^{\frac{1}{2}} \frac{1}{\sqrt{2}}  \sqrt{1+|\chi|+|\chi-1|}\;.
\end{equation}
Since there is a single superconformal cross-ratio of each chirality, it is simple to uplift this result to the full supermultiplet; it suffices to replace $\chi^\pm\to \chi_S^\pm$ and the prefactor of $\left|    \frac{1}{x_{12}x_{34}}
\right|^{\frac{1}{2}}$ with its supersymmetric analog $\frac{1}{|\langle12\rangle\langle34\rangle|^{\frac{1}{2}}}$. The result is thus
\begin{align}\label{eq:conformal-4pt}
    \langle \Phi(x_1)\bar{\Phi}(x_2)\Phi(x_3)\bar{\Phi}(x_4) \rangle = \frac{1}{|\langle12\rangle\langle34\rangle|^{\frac{1}{2}}}\frac{1}{\sqrt{2}}  \sqrt{1+|\chi_s|+|\chi_s-1|}\;.
\end{align}

As a consistency check, we have checked that expanding both sides in the superspace coordinates gives the expected result for four-point functions of other components of $\Phi$.

\subsection{Chaos}

We can now compute the chaos exponent using the procedure outlined in section \ref{sec:chaos}. This requires diagonalizing the retarded kernel. We perform this diagonalization by first writing down the kernel for the standard four-point function, and then performing the analytic continuation required to obtain the retarded kernel. In the following we will focus on the bottom component of all four-point functions, assuming that they produce that leading behavior at long times, as was the case in similar models \cite{Murugan:2017eto,Bulycheva:2018qcp,Berkooz:2021ehv}. 

\subsubsection{The chaos exponent at weak coupling}

We start with the eigenvalues of the standard kernel. The eigenvalues of the bosonic kernel are given by 
\begin{equation}
    k(h,\J)  = \frac{\int d^2 X_3 d^2 \bar{X}_4 K \cdot W}{W}\;,
\end{equation}
where $d^2X=d^2xd^2\theta$ and $d^2\bar X=d^2xd^2\bar\theta$. At leading order in $\J$ the kernel $K$ is given by the leading (super-)diagram in figure \ref{fig:kernel} and $W$ is given by \eqref{eq:W_ansatz}. Plugging in the form of the four-point function \eqref{eq:conformal-4pt} we find\footnote{We are ignoring the subtractions here. They are necessary to compute the eigenvalues $k$, but will not contribute to the eigenvalues of the retarded kernel which are our main interest.}
\begin{equation}
    \int KW=\int d^2 X_3 d^2 \bar{X}_4 \langle\Phi_1\bar{\Phi}_2\Phi_3\bar{\Phi}_4\rangle\frac{1}{|\langle34\rangle|^{3/2-2h}}=
    \frac{1}{\sqrt 2|\langle12\rangle|^{1/2}}
    \int d^2 X_3 d^2 \bar{X}_4
     \frac{ \sqrt{1+|\chi_s|+|\chi_s-1|}}{|\langle34\rangle|^{2-2h}}\;.
\end{equation}
We will focus on the bosonic part of this integral. Using
\begin{equation}
    \begin{split}
        \langle12\rangle^{\pm}=x_{12}^{\pm}\;,\qquad & \langle 34 \rangle^{\pm}=x_{34}^{\pm}-2\theta_3^{\pm}\bar{\theta}^{\pm}_4\;,\\
        \chi_s^{\pm}=\frac{x_{12}^{\pm}(x_{34}^{\pm}-2\theta_3^{\pm}\bar{\theta}^{\pm}_4)}{x_{14}^{\pm}x_{32}^{\pm}}&=\chi^{\pm}(1-\frac{2\theta_3^{\pm}\bar{\theta}^{\pm}_4}{x_{34}^{\pm}})\;.
    \end{split}
\end{equation}
we can perform the Grassman integrals to obtain
\begin{equation}\label{eq:bosonic_kernel}
        k=\frac{1}{|z_{12}|^{1/2}}\int d^2x_3 d^2x_4 \frac{(x_{34}^-)^{h-2}({x}_{34}^+)^{h-2}}{4\sqrt{2}}\mathcal{I}(\chi^\pm)
\end{equation}
where
\begin{equation}
    \begin{split}
        \mathcal{I}=& \frac{1}{\left(|\chi-1|+|\chi|+1\right)^{3/2}}
        \left[\frac{|\chi|\left(4h\left(|\chi-1|-\chi^++1\right)-2|\chi-1|+3\chi^+-2\right)+\chi^-\left(7|\chi|-2\chi^++4\right)}{|\chi-1|}\right.\\
        &+\frac{\chi^-\left(4h\left(2\chi^+|\chi-1|+2\chi^+|\chi|-|\chi-1|-2|\chi|+\chi^+-1\right)-6\chi^+\left(|\chi-1|+|\chi|\right)+4|\chi-1|\right)}{|\chi-1|}\\
        &\left.-4(h-1)\left(|\chi-1|+|\chi|+1\right)\left(\frac{|\chi|-\chi^+\left(|\chi-1|+|\chi|\right)}{\chi^+-1}-4(h-1)\left(|\chi-1|+|\chi|+1\right)\right)\right]\;.
    \end{split}
\end{equation}

We can now use analytic continuation to find the eigenvalues of the retarded kernel at finite temperature. More precisely, first we need to go to the cylinder, the do the analytic continuation. The mapping to the cylinder is given by \cite{Murugan:2017eto}
\begin{equation}
    \begin{array}{lll}
u=e^{x-t}, & v=e^{-x-t}, & \text { (left rail) } \\
u=-e^{x-t}, & v=-e^{-x-t},& \text { (right rail) }
\end{array}
\end{equation}
with the rails identified in figure \ref{fig:time_contour}. The analytic continuation is done by taking the following times:
    \begin{equation}\label{eq:analytic_cont}
      \tau_{1}=\beta/2\pm\epsilon_1+it_{1},\quad \tau_{2}=\pm\epsilon_2+it_{2},\quad 
      \tau_{3}=\beta/2+it_{3},\quad 
      \tau_{4}=it_{4}\;.
    \end{equation}
The procedure is then to compute the four-point function for each of the four choices of signs for the $\epsilon$'s, and then add them up where each term gets a sign which is positive if both epsilons have the same sign and negative otherwise. This gives the double-commutator
\begin{equation}
    \left\langle [\Phi_{i}\left(\beta/2+it_{1}\right),\Phi_{j}\left(\beta/2+it_{3}\right)][\overline{\Phi}_{i}\left(it_{2}\right),\overline{\Phi}_{j}\left(it_{4}\right)]\right\rangle
    \end{equation} which can be written as
\begin{equation}
   \left\langle \left(\Phi_{i}\left(\beta/2+\epsilon_1+it_{1}\right)-\Phi_{i}\left(\beta/2-\epsilon_1+it_{1}\right)\right)\left(\overline{\Phi}_{i}\left(\epsilon_2+it_{2}\right)-\overline{\Phi}_{i}\left(-\epsilon_2+it_{2}\right)\right)\Phi_{j}\left(\beta/2+it_{3}\right)\overline{\Phi}_{j}\left(it_{4}\right)\right\rangle \;,
\end{equation}
where we eventually take the limit $\epsilon_i\to 0$.
Note that for most combinations of $\tau$'s, the $\epsilon$-dependence drops out in this limit. For example, in the combination $\tau_1-\tau_4$ the real part is always dominated by $\beta$ and not by the $\epsilon$ term, and so the sign of $\epsilon$ does not affect the result when we take $\epsilon\to 0$. As a result, $\chi^{\pm}(\epsilon)=\chi^{\pm}$ does not depend on $\epsilon$:
\begin{equation}
    \chi^{\pm}=\frac{\sinh\frac{x_{12}\pm i\tau_{12}}{2}\sinh\frac{x_{34}\pm i\tau_{34}}{2}}{\sinh\frac{x_{14}\pm i\tau_{14}}{2}\sinh\frac{x_{32}\pm i\tau_{32}}{2}}\;.
\end{equation}
However, $|\chi-1|$ does depend on $\epsilon$. Following this procedure for $|\chi-1|$ we find that the analytic continuation takes
\begin{equation}
    |\chi-1|=\sqrt{(\chi^--1)(\chi^+-1)}\to -4\sqrt{|(\chi^--1)(\chi^+-1)|}\;,
\end{equation}
assuming $t_{13}>\left|x_{13}\right|$ and $t_{24}>\left|x_{24}\right|$, and the expression vanishes otherwise.

Up to overall factors which will not be important, the integral then becomes
\begin{equation}
    \begin{split}
        k_R=\int (KW)_R=\int du_3 dv_3 du_4dv_4\frac{1}{u_{34}^{2-h}v_{34}^{2-h}} \mathcal{I}_R
    \end{split}
\end{equation}
where $\mathcal{I}_R$ is given by taking $\mathcal{I}$ and subtracting $\mathcal{I}$ but where each $|\chi-1|$ is replaced by $-|\chi-1|$. In these new variables, we have
\begin{equation}
    \chi=\frac{u_{12}u_{34}}{u_{14}u_{32}},\qquad \bar\chi=\frac{v_{12}v_{34}}{v_{14}v_{32}}\;,
\end{equation}
and the integration region is 
$u_3>u_1,\;u_2>u_4,\;v_3>v_1,\;v_2>v_4$.

As discussed in \ref{sec:chaos}, the chaos exponent as $\J\to 0$ is found by looking for divergences, which should appear at large $|u_i|,|v_i|$; the chaos exponent is $\lambda_0=-2h_*$ where $h_*$ is the value for which the integral diverges. We can find this analytically, since we expect the divergence to come from taking large $u_3,v_3,u_4,v_4$ (and they are all of the same order of magnitude), which means we are taking $\chi,\bar\chi\to 0$. In this limit the integrand behaves as
\begin{equation}
    \frac{1}{u_{34}^{2-h}v_{34}^{2-h}}
\end{equation}
which diverges for $h\geq0$. So the chaos exponent at weak coupling is 
\begin{equation}
    \lambda_L(\J\to 0)=0\;.
\end{equation}

\subsubsection{Contributions from higher $n$-point functions}

As discussed in section \ref{sec:chaos}, we must make sure that contributions to the kernel from higher $n$-point functions don't diverge at lower values of $\lambda$ than $\lambda_L(\J\to 0)=0$, otherwise the approximations made above are invalid. We know all of the higher $n$-point functions since we have mapped the theory to a free theory, and so it is a matter of plugging the results into the kernel.\footnote{As in \cite{Berkooz:2021ehv}, we can ignore the contribution of subtractions in this calculation.}

As an example, we consider the contribution from the bottom component of the $2n$-point function for any $n$ to the kernel in figure \ref{fig:kernel}. Using the free field realization, we find that any $2n$-point function of the bottom component $\phi$ of the superfield $\Phi$ takes the form
\begin{equation}\label{eq:bottom_component}
    \langle\phi(x_1)...\phi(x_n)\bar\phi(y_1)...\bar\phi(y_n)\rangle=\frac{1}{2^{n/2}}\sqrt{\sum_{\substack{\epsilon_{i}^{x}=\pm1,\;\;\epsilon_{i}^{y}=\pm1,\\ \sum\epsilon_{i}^{x}+\epsilon_{i}^{y}=0}}\frac{\prod_{i<j}|x_{ij}|^{\left(\epsilon_{i}^{x}\epsilon_{j}^{x}+1\right)/2}|y_{ij}|^{\left(\epsilon_{i}^{y}\epsilon_{j}^{y}+1\right)/2}}{\prod_{i,j}|x_{i}-y_{j}|^{\left(1-\epsilon_{i}^{x}\epsilon_{j}^{y}\right)/2}}}\;.
\end{equation}
Combined with the propagators \eqref{eq:props}, we can find the contributions to the kernel at any order and look for divergences as $x^+_3\to\infty,x^+_4\to\-\infty$ and similarly for $x_{3,4}^-$. In this limit, the contribution from the six point function behaves as $\frac{1}{|x_{34}|^{4-2h}}$, which matches the behavior found in the contribution from the four-point function. The divergence is thus also at $h=0$, and so our approximation is consistent at this level. One can repeat the analysis for other components of the 2n-point function which contribute to $k_R$, and as a result the approximations discussed above are consistent.

\subsubsection{Continuity conjecture}

We now compute $\lambda_L^0$ and compare to $\lambda_L(\J\to 0)$ in order to show that the continuity conjecture is obeyed. This amounts to taking the bosonic part of the analytically-continued four-point function of a single core CFT \eqref{eq:conformal-4pt} and studying its behavior in the large-time limit $t_3\sim t_4=t\to \infty$.

First we must do the analytic continuation discussed above. The bosonic part of the 4-pt function is
\begin{equation}
    \frac{1}{\sqrt{2}}\frac{\sqrt{1+|\chi|+|\chi-1|}}{|z_{12}z_{34}|^{\frac{1}{2}}}\;.
\end{equation}
Mapping to the cylinder and performing the analytic continuation discussed above, we find 
\begin{equation}
    \frac{\sqrt{1+|\chi|+|\chi-1|}}{|z_{12}z_{34}|^{\frac{1}{2}}}\to 2\theta(t_{13}-|x_{13}|)\theta(t_{24}-|x_{24}|)\frac{\sqrt{1+|\chi|+|\chi-1|}-\sqrt{1+|\chi|-|\chi-1|}}{|z_{12}z_{34}|^{\frac{1}{2}}}\;.
\end{equation}
Now we can take the large-time limit, which amounts to taking $\chi,\bar{\chi}\sim e^{-t}$ for $t$ large. We find
\begin{equation}
    \sqrt{1+|\chi|+|\chi-1|}-\sqrt{1+|\chi|-|\chi-1|}\approx \sqrt 2 -\sqrt 2 e^{-t/2}+...
\end{equation}
and so the analytically-continued four-point function is finite in the large-$t$ limit, since it is not exponentially vanishing in $t$. As a result, $\lambda_L^0=0$ since there is no exponential behavior. We thus find that $\lambda_L^0=\lambda(\J\to 0)=0$ and so the continuity conjecture is obeyed.

\section{Conclusions}\label{sec:conclusions}

In this paper we have proven the continuity conjecture in QM and given additional evidence for it in $2d$. We also studied the disordered $\mathcal{A}_3$ minimal model at leading order in the random coupling $\J$.

There are many interesting questions left for future work. For example, in both of the $2d$ examples given here (the chiral SYK model and the disordered $\mathcal{A}_3$ model), the chaos exponent at small coupling was $\lambda_L^0=0$. In \cite{Berkooz:2021ehv}, the example of the disordered $\mathcal{A}_2$ minimal model was also shown to have $\lambda_L^0=0$. These examples should be compared to the case of disordered generalized free fields where $\lambda_L^0<0$, leading to a discontinuous transition into chaos as $\J$ increases \cite{Berkooz:2021ehv}.\footnote{This behavior is not related to the non-locality of the generalized free field theory; for example, a negative $\lambda_L^0$ was found in \cite{Caron-Huot:2017vep}, although this corresponds to the chaos exponent on thermal Rindler space and not flat space.} We can ask why this interesting behavior did not occur here, and an obvious conjecture is that $\lambda_L^0$ can only be non-zero when the core CFT does not have a free theory realization. Unfortunately, this means that performing computations in theories with $\lambda_L^0<0$ will be difficult, since the computations require knowledge of exact $n$-point functions of the CFT. One possible direction would be to perform these computations in $\mathcal{N}=2$ minimal models with higher central charges, where the four-point functions are known exactly, but performing the necessary integrals over them is a very complicated technical task which we leave to future work.

Another future direction would be to prove the continuity conjecture in $2d$. This becomes more complicated due to the appearance of the butterfly velocity. The divergence which leads to the chaos exponent at weak coupling appears at large times, but can appear at any value of the ratio $v=t/x$, and an additional integral must be performed over these values. The final result depends nontrivially on $v$, and the result must be studied carefully in order for the chaos exponent to be read off. However, the chiral example discussed above is evidence that even with a nontrivial butterfly velocity, the continuity conjecture is obeyed. It is yet to be seen whether it is obeyed only for the velocity which leads to the maximal chaos exponent, or whether it is obeyed for any velocity.

It would also be interesting to extend calculations of $\lambda_L(\J)$ for a continuous range of $\J$ to higher dimensions, building on \cite{Chang:2021fmd}. For example, considering $N$ free bosons in $3d$, the deformation $((\varphi_i)^2)^3$ is exactly marginal at leading order in $1/N$ \cite{PhysRevLett.52.1188}. Similarly for $N$ free matter multiplets in a $3d$ $\mathcal{N}=1$ supersymmetric theory, the superpotential deformation $(|\Phi_i|^2)^2$ is also exactly marginal at leading order in $1/N$ \cite{Aharony:2019mbc}. One could perform a computation of the chaos exponent analogous to that of the $O(N)$ model \cite{Chowdhury:2017jzb} as a function of the exactly marginal deformation.\footnote{A better understanding of whether it is enough to be exactly marginal at leading order in $1/N$ is required in order to perform these computations.} The chaos exponent is expected to be non-positive since the theory is close to being free, and it would be interesting to see if it is exactly zero.

\section*{Acknowledgements}

The authors would like to thank E. Y. Urbach and N. Silberstein for collaborating on this work in its early stages, and B. Lian for collaboration on related topics. The authors would also like to thank Micha Berkooz and Doron Gepner for enlightening discussions.
The work of RRK was supported in part by an Israel Science Foundation (ISF) center for excellence grant (grant number 2289/18), by ISF grant no. 2159/22, by Simons Foundation grant 994296 (Simons Collaboration on Confinement and QCD Strings), by grant no. 2018068 from the United States-Israel Binational Science Foundation (BSF), by the Minerva foundation with funding from the Federal German Ministry for Education and Research, by the German Research Foundation through a German-Israeli Project Cooperation (DIP) grant ``Holography and the Swampland'', and by a research grant from Martin Eisenstein.

\newpage

\begin{appendix}

\addtocontents{toc}{\protect\setcounter{tocdepth}{1}}

\section{ Superappendix}\label{app:superappendix}

\subsection{The supersymmetry algebra}

$\mathcal{N} = (2,2) $ SUSY in two dimensions is closely related to $\mathcal{N} = 1$ in four dimensions. There are two real spinors' worth of supercharges: $Q^{(a)} _{\alpha} $, where $\alpha $ is a Lorentz (spinor) index and $a = 1 , ... , \mathcal{N} $ is a separate (R-symmetry) index. The SUSY algebra is
\begin{align}
\{ Q^{a} _{\alpha } ,  Q^{b} _{\beta } \} & = 2\gamma ^{\mu} _{\alpha \beta} P_{\mu} \delta ^{ab}\;,
\end{align}
and the gamma matrices are:
\begin{align*}
    \gamma ^0 _{\alpha \beta } = \begin{bmatrix}
                                    1 & 0 \\
                                    0 & -1
                                  \end{bmatrix}\;,
                        \qquad
    \gamma ^1 _{\alpha \beta } = \begin{bmatrix}
                                    1 & 0 \\
                                    0 & 1
                                  \end{bmatrix}\;.
\end{align*}                    
We can break the spinor representation into the left/right-moving sectors, which are one-dimensional and transform by a phase under the Lorentz group:
\begin{align}
    Q ^{a}_{\pm} & \rightarrow e^{ \pm \frac{1}{2} \eta} Q^{a}_{\pm} 
\end{align}
In Lorentzian signature, the supercharges are real:
\begin{align}
    \overline{ Q ^a _{\pm} }  & = \lb Q^{a}_{ \pm} \rb ^* = Q^{a}_{ \pm} 
    \label{eqn:conj}
\end{align}

\subsubsection{ $\mathcal{N} = (1,1)$ }

A real $\mathcal{N} = (1,1) $ scalar superfield $\mathcal{A} (x, \theta ) $ consists of a real scalar $\phi (x)$ and its superpartners: a Majorana spinor $\psi _{\alpha} (x)$, and another scalar $F(x)$.
\begin{align}
    \mathcal{A} (x, \theta ) & = e^{ \theta ^{\alpha } Q_{\alpha } } \phi (x) \label{eqn:supf} \\
    & = \phi (x) + \theta ^{\alpha} \psi_{\alpha } (x) + \theta ^{\alpha} \theta ^{\beta} \epsilon _{\alpha \beta} F (x)  + \dots \nonumber \\
         & = \phi (x) + \theta ^{+} \psi_+ (x) + \theta ^{-} \psi_- (x) + \theta ^{+} \theta ^{-}  F (x) +\dots
\end{align}
where $\theta ^{\alpha } $ is a real Grassman spinor,\footnote{$\lb \theta ^{ \pm } \rb ^2 = 0 $.} and is given a Lorentz transformation so that the superfield is a scalar. We've broken up $\psi_{\alpha} (x)$ into Lorentz irreps, and the equations of motion will tell us that $\psi_{\pm}$ are left/right moving (holomorphic/antiholomorphic in Euclidean signature). Higher components are descendants since the supercharge squares to a derivative.
There is a discrete R-symmetry
\begin{align}
    Q_{\alpha} & \rightarrow -Q_{\alpha}\;,
\end{align}
that leaves the SUSY algebra invariant. This gives the following transformations for the fields:
\begin{align}
    \phi (x) & \rightarrow (-1) ^{\gamma } \phi (x)\;, \nonumber \\
    \psi _{\alpha} (x) & \rightarrow (-1) ^{\gamma + 1 } \psi _{\alpha}  (x)\;, \nonumber \\
    F(x) & \rightarrow (-1) ^{\gamma + 2 } F (x) \;.
\end{align}
With $ \theta ^{\alpha } \rightarrow - \theta ^{\alpha } $, the superfield $ \mathcal{A} (x, \theta ) $ has the same R-charge as its bottom component (here, the scalar; see \eqref{eqn:supf}).
The Lagrangian (free-field plus superpotential) in terms of superfields may be written as
\begin{align}
\mathcal{L} & = \int d^2 \theta (  \frac{1}{2} \mathcal{D} \mathcal{A} \bar{ \mathcal{D} } \mathcal{A}+ W ( \mathcal{A} ) )\;.
\end{align}

\subsubsection{ $\mathcal{N} = (2,2) $}

The $\mathcal{N} = (2,2)$ algebra has two Majorana supercharges, and we need two real Grassman spinors $\theta ^{\alpha} _a$. A complex superfield $\mathcal{X}$ takes the form
\begin{align}
    \mathcal{X} (x, \theta ^a ) & =  e^{ \theta _a ^{\alpha} Q ^a _{\alpha }} \phi(x ) 
      = \phi  + \theta _a ^{\alpha} \psi ^a _{\alpha } + ... \nonumber \\
         & = \phi (x) + \theta ^{+} _a \psi ^a _+ (x) + \theta ^{-} _a \psi ^a _- (x) + ... \nonumber \\
    \bar{ \mathcal{X} } (x, \theta ^a ) & = \bar{ \phi }  + \theta _a ^{\alpha} \bar{ \psi } ^a _{\alpha } + ... \nonumber \\
         & = \bar{ \phi } (x) + \theta ^{+} _a \bar{ \psi } ^a _+ (x) + \theta ^{-} _ a \bar{ \psi } ^a _- (x) + ...
\end{align}
The bar stands for complex conjugation (see \eqref{eqn:conj}). It is convenient to use complex combinations of the supercharges:
\begin{align}
    Q _{\pm} = Q ^1 _{\pm} + i Q ^2 _{\pm}\;, \qquad
    \overline{Q} _{\pm} = Q ^1 _{\pm} - i Q ^2 _{\pm}\;, \qquad
    \overline{ ( Q _{\pm} )} = \overline{ Q } _{\pm}\;.
\end{align}
The $ \mathcal{N} = (2,2) $ SUSY algebra may be written as:
\begin{align}
    \{ Q_{\pm} , \overline{Q} _{\pm} \}  & = 2 P_{\pm } = 2 ( \pm P_0 + P_1 ) \;.
\end{align}
The Grassman variables are now:
\begin{align}
    \theta ^{\pm} = \theta _1 ^{\pm} + i \theta _2 ^{\pm}\;, \qquad
    \overline{\theta} ^{\pm} = \theta _1 ^{\pm} - i \theta _2 ^{\pm}\;.
\end{align}
The superfield takes the form
\begin{align}
    \mathcal{X} (x , \theta , \bar{\theta } ) & = e^{  \bar{ \theta } ^{\alpha} Q_{\alpha} +  \theta  ^{\alpha } \bar{Q}_{\alpha } } \phi (x) \label{eqn:supf2} \\
    & = 
     \phi (x) + \bar{ \theta } ^{+} \tilde{\psi } _+ (x) +  \theta ^{+}  \psi   _+ (x) + \bar{ \theta } ^{-} \tilde{\psi } _- (x) + \theta ^{-} \psi  _- (x) + ... 
     \nonumber \\
     \bar{\mathcal{X} } (x , \theta , \bar{\theta } )  
    & = 
     \bar{ \phi } (x) + \bar{\theta } ^{+} \bar{ \psi  } _+ (x) + \theta ^{+} \bar{ \tilde{ \psi }} _+ (x) + \bar{\theta }^{-} \bar{ \psi } _- (x) + \theta ^{-}  \bar{\tilde{ \psi } } (x) + ... 
\end{align}
with
\begin{align}
    \psi _{\pm} = \psi _{\pm }^1 - i \psi _{\pm }^2\;, \qquad
    \tilde{ \psi } _{\pm}  = \psi _{\pm }^1 + i  \psi _{\pm }^2 \;.
\end{align}
The $\mathfrak{u}(1)_R$ symmetry rotates $(Q, \bar{Q})$ in opposite directions:
\begin{align}
   & Q_{\pm} \rightarrow e^{ i \alpha } Q_{\pm}\;,  \qquad
    \bar{Q}_{\pm} \rightarrow e^{ - i \alpha } \bar{Q}_{\pm} \;,
        \nonumber \\
   & \theta ^{\pm} \rightarrow e^{i \alpha } \theta ^{\pm}\;, \qquad
    \bar{ \theta } ^{\pm} \rightarrow e^{- i \alpha } \bar{ \theta } ^{\pm}\;.
\end{align}
There is also an axial $\mathfrak{u}(1)_{\tilde{R}}$ which rotates the $\pm$ components with opposite phases.

\paragraph{Chirality}

The superfield \eqref{eqn:supf2} is in a reducible representation. Imposing
\begin{align}
    Q ( \phi (x) ) &  = \bar{Q} ( \bar{\phi} (x) ) = 0\;, \nonumber
\end{align}
we have
\begin{align}
    \Phi (x , \theta ,  \bar{ \theta } ) & = e^{ ( \bar{ \theta } ^{\alpha} Q_{\alpha} + \theta ^{\alpha } \bar{Q}_{\alpha } ) } \phi (x) =
     e^{ ( \theta  ^{\alpha} \bar{Q}_{\alpha} + i \theta  ^{\alpha } \gamma ^{\mu } \bar{ \theta } P_{\mu}) } \cancel{e^{ \bar{ \theta } ^{\alpha} Q_{\alpha }  } } \phi (x)
     \nonumber \\
     &  = \phi (x) + \theta  ^{+} \psi _+ (x) + \theta  ^{-} \psi _- (x)  + \theta  ^{+} \theta  ^{-} F(x) + ... 
\nonumber \\
     \bar{\Phi } (x , \theta, \bar{ \theta } )  
     & = 
     \bar{\phi } (x) + \bar{ \theta } ^{+} \bar{\psi } _+ (x) + \bar{ \theta } ^{-} \bar{\psi } _- (x) + \bar{\theta } ^{+} \bar{\theta } ^{-} \bar{F}(x) + ...
\end{align}
This makes $\Phi (x, \theta ,\bar{\theta } )$ a chiral superfield. An $\mathcal{N} = (2,2)$ chiral (scalar) multiplet contains only one complex fermion.
It is easy to write down interactions for such superfields:
\begin{align}
    \mathcal{L} & = \int d ^2 \theta d^2 \bar{\theta } \bar{\Phi }  \Phi + \int d^2 \theta W(\Phi ) + \int d^2 \bar{\theta} \bar{W} ( \bar{\theta }) \;.
\end{align}
Of course, these are not the only possible interactions.

\subsection{The $ \mathcal{A}_3 $ minimal model}

We are interested in the following $ \mathcal{N} = (2,2) $ theory:
\begin{align}
\mathcal{L} & = \int d^ 2 \theta d^2 \bar{\theta } \bar{ X} X + \int d^2 \theta X ^4+ \int d^2 \bar{\theta } \bar{X }^4  \;,
\end{align}
with $X$ a chiral superfield.

\begin{table}
\centering
\caption{Operators from the Ising model. The last entry $\chi $ is the fermion. $\sigma $ and $\mu $ are the spin and disorder operators; these aren't mutually local.}
\label{tab:Ising}
\begin{tabular}{@{}clllc@{}}\toprule
            $\mathcal{O} $   &    $ \Delta $  & $\ell$   \\
            \cmidrule{1-3}
            $ \sigma $ & $ \frac{1}{8} $ & 0 \\
            $ \mu $ & $ \frac{1}{8} $ & 0 \\
            $ \epsilon $ & $ 1 $ & 0 \\
            $ \chi _{\pm} $ & $\tfrac{1}{2} $ & $\tfrac{1}{2}$ \\
            \bottomrule
        \end{tabular}
\end{table} 
It's easily seen that the interactions preserve $\mathfrak{u}(1)_R$ with $q_{X} = 1/2$; this blocks the superpotential from picking up other terms under RG flow. The theory in the IR is a Landau-Ginsburg minimal model with central charge $c = \tfrac{3}{2}$ that may also be obtained by putting together a free scalar and the Ising model. In the following, $H$ is a free boson, and Table \ref{tab:Ising} are the operators we need from the Ising model.

Firstly, we identify the SUSY generators\footnote{Recall that the SUSY currents are operators with dimensions $ (\tfrac{3}{2},0) $ or $ (0, \tfrac{3}{2}) $ ($G_{\pm} $ respectively).} and R symmetry current:
\begin{align}
    G _{ \pm } & = \chi _{\pm } \exp \lb i \sqrt{2} H_{\pm } \rb\;,
    \nonumber \\
    \bar{G} _{ \pm } & = \chi _{\pm } \exp \lb - i \sqrt{2} H_{\pm } \rb\;,
    \nonumber \\
    j_{\pm} ^{(R)} & = \frac{i }{\sqrt{2} } \partial _{\pm} H\;.
\label{eqn:curr}
\end{align}
$H_{\pm} $ are the left/right moving parts of the scalar field $H$. The bottom components of the superconformal primaries are given by:
\begin{align}
    \phi = \sigma \exp \lb i \frac{H}{ 2 \sqrt{2} } \rb\;,  & \qquad 
        \bar{\phi } = \sigma \exp \lb - i \frac{H}{ 2 \sqrt{2} } \rb\;,
    \nonumber \\
    \phi ^2 = \exp \lb i \frac{H}{ \sqrt{2} } \rb\;,  & \qquad 
        \bar{\phi } ^2 = \exp \lb - i \frac{H}{ \sqrt{2} } \rb\;,
    \nonumber \\
    \bar{\phi} \phi & = \epsilon\;.
    \label{eqn:A3primaries}
\end{align}
We can then find the free field expressions for their SUSY descendants using \eqref{eqn:curr} and the OPE's:
\begin{align}
    \chi _{\pm } (x^{\pm} ) \times \sigma (x)  \sim
        \frac{1}{ \sqrt{ x^{\pm} }} \mu\;,
        \qquad
    \chi _{\pm } (x^{\pm} ) \times \mu (x)  \sim
        \frac{1}{ \sqrt{ x^{\pm} }} \sigma\;.
\end{align}
This leads to:
\begin{align}
    \bar{G}_{+}  (x^+) \times 
    \phi &
    \sim
    \frac{ 1 }{ x^+ } \ \mu \exp \lb \frac{i }{2 \sqrt{2} } \lb - 3 H_+ 
    + H_- \rb \rb  = \frac{1}{ x^{ +} } \psi _{+ } \;,
    \nonumber \\
    \bar{G}_{-} (x^-) \times 
    \phi 
    & \sim 
    \frac{ 1 }{ x^- } \ \mu \exp \lb \frac{i }{2 \sqrt{2} } \lb  H_+ 
    -3 H_- \rb \rb  = \frac{1}{ x^{ - } } \psi _{- } \;,
    \nonumber \\ 
    \bar{G}_{+} (x^+) \times 
    \psi _- 
    & \sim 
    \frac{ 1 }{ x^+ } \ \sigma \exp \lb -i \frac{3 H }{2 \sqrt{2} } \rb  = \frac{1}{ x^{ + } } F\;,
    \nonumber \\ 
    \bar{G}_{-} (x^-) \times 
    \psi _+ 
    & \sim 
    \frac{ 1 }{ x^- } \ \sigma \exp \lb -i \frac{3 H }{2 \sqrt{2} } \rb  = \frac{1}{ x^{ - } } F\;.
    \label{eqn:SUSYdesc}
\end{align}
We summarise these results in Table \ref{tab:map}.
\begin{table}
\centering
\ra{1.5}
\caption{The operators in the multiplet of $ X $ in the $\mathcal{A}_3 $ minimal model.}
\label{tab:map}
\begin{tabular}{@{}cllc@{}}
\toprule
            $\mathcal{O}$ & $\mathcal{O}_{\text{free}} $   &    $ ( h ,\bar{h} ) $ & $q_R $  \\
            \cmidrule{1-4}
            $ \phi $ & $ \sigma \ e^{i \frac{H}{2 \sqrt{2} } }  $ & $ \lb \frac{1}{8} , \frac{1}{8} \rb $ & $ \lb \frac{1}{4} , \frac{1}{4} \rb $ \\
            $ \psi _{+} $ & $ \mu \ e^{i \frac{-3 H_+ + H_{-} }{2 \sqrt{2} } }  $ & $ \lb \frac{5}{8} , \frac{1}{8}\rb $ & $ \lb -\frac{3}{4} , \frac{1}{4} \rb $ \\
            $ \psi _{-} $ & $ \mu \ e^{i \frac{ H_+ - 3H_{-} }{2 \sqrt{2} } }  $ & $ \lb \frac{1}{8} ,\frac{5}{8} \rb $ & $ \lb \frac{1}{4} , -\frac{3}{4} \rb $\\
            $ F $ & $ \sigma \ e^{- i \frac{3 H}{2 \sqrt{2} } }  $ &  $ \lb \frac{5}{8} ,\frac{5}{8} \rb $ & $ \lb -\frac{3}{4} , -\frac{3}{4} \rb $ \\
            \bottomrule
        \end{tabular}
\end{table}

\paragraph{Mutual locality of the operators}

Firstly, $ G $, $ \bar{G} $, and $ j^{R}$ are mutually local (of course, the fermionic operators are mutually local only in the fermionic sense -- they pick up a $ -1 $) since the bosonic pieces are the conserved currents of the $\mathfrak{su}(2)_1 $ model. Now, all the operators within the $ X $ and $ \bar{ X } $ multiplet are obtained by operator products with the currents. So to show that these are mutually local, we just need to show that the bottom components don't see branch cuts with respect to each other or the currents.

The OPE of two vertex operators (analytically continued to Euclidean space) take the form:
\begin{align}
    e^{ i k_1 . H } (z) \times e^{ i k_2 . H } (0) & \sim
    z ^{k_1 ^R k_2 ^R } \bar{z} ^{ k_1 ^L k_2 ^L }  e^{ i (k_1 + k_2 ) . H } (0)
    \nonumber \\
    & =
    ( z \bar{z} )^{ \tfrac{1}{2} k_1 . k_2  } \lb \frac{ z}{ \bar{z} } \rb ^{ -\tfrac{1}{2}( k_1 ^L k_2 ^L - k_1 ^R k_2 ^R ) } e^{ i (k_1 + k_2 ) . H } (0)
\end{align}
where $
    k_{1,2} . H = k_{1,2}^R H_+ +  k_{1,2}^L H_-
$ and $k_1 . k_2 = k_1^R k_2^R + k_1^L k_2^L $. When one operator goes around the other, the right-hand side picks up $ e^{  2\pi i ( k_1 ^L k_2 ^L - k_1 ^R k_2 ^R ) }$. Setting this phase to 1 gives the Narain condition:
\begin{align}
    k_1 \odot k_2 = k_1 ^L k_2 ^L - k_1 ^R k_2 ^R \in \mathbbm{Z}
    \label{eqn:Narain}
\end{align}
This will be used to check locality between vertex operators. The R-current plays well with all vertex operators, so we won't have to bother with it.

Coming back to our operators, the Ising parts of $\phi $ and $\bar{\phi }$ are mutually local. So we just need to examine the vertex operators, which are easily seen to satisfy \eqref{eqn:Narain}. Actually it quite easy to see that all the bottom components listed in \eqref{eqn:A3primaries} are mutually local. The first nontrivial check is between $(G, \bar{G})$ and $\phi $. But we can easily verify from \eqref{eqn:SUSYdesc} that there are no branch cuts; the same conclusion holds for $(G, \bar{G})$ and $\bar{\phi } $ via the conjugate of \eqref{eqn:SUSYdesc}.
$ \bar{\phi } \phi $ and the currents are also easily kosher. To work out $ \phi ^2 $ case, we need only focus on the vertex operators. The only non-zero $\odot $ products are $ (\pm \sqrt{2} , 0 ) \odot ( \tfrac{1}{\sqrt{2} } , 0 ) = \pm 1 $.

\section{Simplification of \texorpdfstring{$1d$}{1d} integral}\label{app:1d_integral}

Using the iterative ladder structure, we can write the full retarded kernel as
\begin{equation}
    K_{R}(1,2)=K_{0}(1,2;3,4)G_{lr}^{q-2}(3,4)\;.
\end{equation}
At leading order, $K_{0}$ is just the subtracted 4-pt function. It is more convenient to work with the rescaled expression
\begin{equation}
    K_R'=e^{\Delta\left(t_1 +t_2 -t_3 -t_4 \right)}K_{R}\;.
\end{equation}
Next, we perform the change of variables $z=e^{-t}$ on the left rail (points 1,3) and $z=-e^{-t}$ on the right (points 2,4), see figure \ref{fig:time_contour}, leading to
\begin{equation}
    K_R'dt_3 dt_4 =K_R'\frac{dz_3 }{z_3 }\frac{dz_4 }{z_4 }\;.
\end{equation}
The two-points are fixed up to a constant that we will ignore:
\begin{equation}
    \begin{split}
        G_{lr}	&=\frac{1}{\left(2\cosh\frac{t_3 -t_4 }{2}\right)^{2\Delta}}=\frac{z_3 ^{\Delta}z_4 ^{\Delta}}{(z_{34})^{2\Delta}}\;,\\
        G_{R}\left(1,3\right)&=\theta\left(t_1 -t_3 \right)\frac{2\cos(\pi\Delta)b}{\left(2\sinh\frac{t_1 -t_3 }{2}\right)^{2\Delta}}=\theta(z_3-z_1)\frac{2\cos(\pi\Delta)bz_1 ^{\Delta}z_3 ^{\Delta}}{(z_{13})^{2\Delta}}\;.
    \end{split}
\end{equation}

Now we use the fact that $K_{0}$ is a (sum of) four-point functions, so we can write it as
\begin{equation}
    K_{0}^{\prime}=e^{\Delta\left(t_1 +t_2 -t_3 -t_4 \right)}G_{lr}(1,2)G_{lr}(3,4)\mathcal{G}(\chi)
\end{equation}
with $\chi$ the conformal cross-ratio and $G$ some function. Putting this together and setting $\Delta=1/q$ we find
\begin{equation}
    K_R'dt_3 dt_4 =\frac{\mathcal{G}_R(\chi)dz_3 dz_4 }{(z_{12})^{2\Delta}(z_{34})^{2\Delta\left(q-1\right)}}
\end{equation}
Finally, applying $K_R'$ to the eigenfunction  $W_{+}'=\frac{1}{|z_{34}|^{2\Delta+\lambda}}$ we find
\begin{equation}
K_R'W_{+}'dt_3 dt_4 =\frac{\mathcal{G}_R(\chi)dz_3 dz_4 }{|z_{12}|^{2\Delta}|z_{34}|^{2+\lambda}}\;.
\end{equation}

\end{appendix}

\printbibliography

\end{document}